\documentclass[11pt]{article}
\usepackage[a4paper,margin=0.7in]{geometry}
\usepackage[english]{babel}
\usepackage{xr}

\usepackage{setspace}
\usepackage{tikz}
\usetikzlibrary{arrows,automata,positioning}
\usepackage{pgfplots}
\pgfplotsset{compat=1.7}

\usepackage{mathrsfs}%, algorithm}%, algorithmic}
\usepackage{url}
\usepackage{hyperref}
\usepackage{breakurl}
\hypersetup{breaklinks=true}
\usepackage[]{amsmath}
\usepackage{amsthm}
\usepackage{fix-cm}
\usepackage[]{amssymb}
\usepackage[]{latexsym}
\usepackage[right]{eurosym}
\usepackage[T1]{fontenc}
\usepackage[]{graphicx}
\usepackage[]{epsfig}
\usepackage{fancyhdr}
\usepackage{multirow}
\usepackage{subcaption}
\usepackage{wrapfig}

\allowdisplaybreaks

\makeatletter

\newcounter{modcount}
\newcommand{\modulo}[2]{%
\setcounter{modcount}{#1}\relax
\ifnum\value{modcount}<#2\relax
\else\relax
\addtocounter{modcount}{-#2}\relax
\modulo{\value{modcount}}{#2}\relax
\fi}
\newcommand{\tablepictures}[4][c]{\begin{tabular}[#1]{@{}c@{}}#2\vspace{0.5cm}\\(\alph{#4}) #3\end{tabular}}
\newcounter{gridsearch}
\newcommand{\tabpic}[2]{
    \stepcounter{gridsearch}
    \modulo{\thegridsearch}{2}
    \ifnum\value{modcount}=0
        \tablepictures[t]{#1}{#2}{gridsearch}\\[2.0cm]
    \else
        \tablepictures[t]{#1}{#2}{gridsearch}&~&
    \fi
}

\makeatother
\newtheorem{lemma}{Lemma}[section]

\newtheorem{example1}[lemma]{Example}
\newtheorem{rem1}[lemma]{Remark}

\newtheorem{alg1}[lemma]{Algorithm}
\newtheorem{me1}[lemma]{Mechanism}

\usepackage{color}
\newcommand{\T}{\top}
\newcommand{\diag}{\operatorname{diag}}

\newcommand\ind[1]{\mathbb{I}_{\{#1\}}}

\begin{document}

%\title{Double Tap: Financial and Economic Analysis of a Zombie Outbreak}
%\title{Brain Drain: Financial and Economic Impacts of a Zombie Outbreak}
%\title{The Undead Hand of the Market: Financial and Economic Impacts of a Zombie Outbreak}
%\title{True Voodoo Economics: Financial and Economic Impacts of a Zombie Outbreak}
\title{Reanimating a Dead Economy: Financial and Economic Analysis of a Zombie Outbreak}
%\title{Necronomics: Financial and Economic Impacts of a Zombie Outbreak}
\author{Zachary Feinstein \thanks{Stevens Institute of Technology, School of Business, Hoboken, NJ 07030, USA, {\tt zfeinste@stevens.edu}}}
\date{\today}
\maketitle
%\addtocounter{footnote}{1}
\abstract{
In this paper, we study the financial and economic implications of a zombie epidemic on a major industrialized nation.  We begin with a consideration of the epidemiological modeling of the zombie contagion.  The emphasis of this work is on the computation of direct and indirect financial consequences of this contagion of the walking dead.  A moderate zombie outbreak leaving 1 \emph{million} people dead in a major industrialized nation could result in GDP losses of 23.44\% over the subsequent year and a drop in financial market of 29.30\%.  We conclude by recommending policy actions necessary to prevent this potential economic collapse.\footnote{Names, characters, businesses, places, events, locales, and incidents are either the products of the author's imagination or used in a fictitious manner. Any resemblance to actual persons, living, dead, or undead, or actual events is purely coincidental.}\\[1em]
}

\section{Introduction}\label{sec:intro}
\emph{Ophiocordyceps mortem}, colloquially and henceforth named the zombie plague, has proliferated throughout history with varying strains of the infection.  The author considers the first modern outbreak to have occurred in 1968 during the so-called \emph{Night of the Living Dead}~\cite{nightlivingdead}, though written records go at least as far back as the early 19th century~\cite{prideprejudice}.  Since that time, outbreaks of the zombie plague have occurred with some regularity.

While epidemiologists~\cite{munz2009zombies,alemi2015zombies} have studied the spread of the zombie plague through unprepared populations, the author is not aware of any such study on the financial and economic implications.  And to paraphrase President Harry S.\ Truman: it's a zombie recession when zombies eat your neighbors; it's a zombie depression when zombies eat you.  As such, the goals of this work are to model the impacts of a new outbreak of the zombie plague on the economic well-being of a major industrialized nation.  Due to the wide-spread panic that may result from the spread of this epidemic, we focus on financial systemic risk triggered by the freezing of short-term lending markets and margin calls, i.e., fire sales and price-mediated contagion as studied in, e.g., \cite{CFS05,GLT15,AFM16,feinstein2015illiquid,feinstein2016leverage,CS17,DE18,BW19,CS19,BF19,feinstein2019leverage}.

The analysis of this paper is as follows.  In Section~\ref{sec:model}, we construct and calibrate an epidemiological model of a zombie outbreak.  In Section~\ref{sec:impacts}, we consider the distribution of financial and economic losses following an outbreak of the zombie plague.  In Section~\ref{sec:results}, we use these results to construct simple policy recommendations to mitigate the economic risks associated with the zombie plague. Section~\ref{sec:conclusion} concludes.  A glossary of symbols utilized in this paper can be found in the appendix.

\section{Epidemiological Model}\label{sec:model}
\begin{figure}[t]
\centering
\begin{tikzpicture}
\node[circle,draw] at (-6cm,0cm) (S) {S};
\node[circle,draw] at (-6cm,-2.5cm) (I) {I};
\node[circle,draw] at (-2cm,0cm) (E) {E};
\node[circle,draw] at (2cm,0cm) (Z) {Z};
\node[circle,draw] at (2cm,-2.5cm) (Q) {Q};
\node[circle,draw] at (6cm,0cm) (R) {R};

\draw[every loop, auto=left, line width=0.6mm, >=latex]
(S) edge[bend right=30] node[left] {$\iota SZ$} (I)
(I) edge[bend right=30] node[right] {$\kappa I$} (S)
(S) edge node {$\epsilon SZ$} (E)
(E) edge node {$\zeta_E E$} (Z)
(E) edge node[left] {$\omega E$} (Q)
(E) edge[bend left=30] node {$\rho_E SE$} (R)
(Z) edge node[above] {$\rho_Z (S+I)Z$} (R)
(Z) edge node[below] {$+\delta Z$} (R)
(Q) edge node {$\zeta_Q Q$} (Z)
(Q) edge node[right] {$\rho_Q SQ + \delta Q$} (R);
\end{tikzpicture}
\caption{Graphical representation of the SZR model utilized in this work.}
\label{fig:szr}
\end{figure}
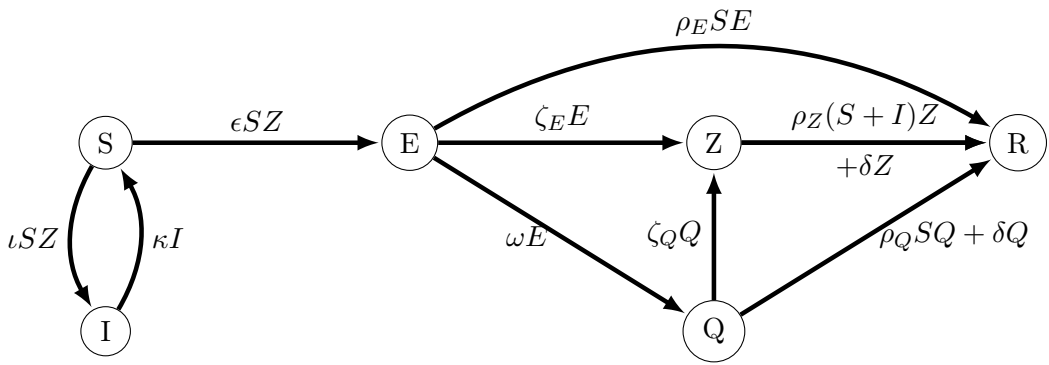

The spread of the zombie plague throughout a population has been a widely studied phenomenon in epidemiological modeling, see, e.g., \cite{munz2009zombies,alemi2015zombies}.  As such, we are able to consider this epidemic using extensions of the well-known susceptible-zombie-removed [SZR] model.  For the purposes of this work, we consider extensions that include an exposure period before exhibiting symptoms of the zombie plague, self-isolation of uninfected individuals, and the quarantining and targeted removal of exposed patients.  In this epidemiological framework there are 6 states that a person can be:
\begin{enumerate}
\item an individual is \textbf{susceptible [S]} if they have not been exposed to the zombie plague but are roaming freely, capable of being infected by the zombified population;
\item an individual is \textbf{isolated [I]} if they have sealed themselves off from civilization (possibly in a small group within an isolated and abandoned mall, as attempted during the 1978 and 2004 flare-ups of the zombie plague~\cite{dawnofdead1,dawnofdead2}) so as to avoid contracting the zombie plague;
\item an individual is \textbf{exposed [E]} if they have been bitten or otherwise been contaminated by a zombified individual;
\item an individual is \textbf{quarantined [Q]} if they have been exposed to the zombie plague but confined to quarters so as to prevent exposing the susceptible population;
\item an individual is \textbf{zombified [Z]} if they are exhibiting all symptoms of the zombie plague and roaming freely, capable of infecting the susceptible population;
\item an individual is \textbf{removed [R]} from the system when they are subject to a double tap\footnote{``In those moments when you're not quite sure if the undead are really dead, dead, don't get all stingy with your bullets,'' Columbus~\cite{zombieland}.}$^{,}$\footnote{Contrary to popular belief, using plants to shoot zombies is not effective.} and thus no longer a threat to the general populace.
\end{enumerate}

As summarized in Figure~\ref{fig:szr}, individuals can, and do, transition between states due to isolation, exposure, quarantining, zombification, and removal at different rates dependent on the sizes of the populations and speed of progression denoted by the Greek letters.  We wish to note that $\zeta_Q \leq \zeta_E$ as the exposed but quarantined population will need to both zombify and escape the quarantine zone in order to be classified as a zombified individual.
Mathematically, we define our SZR model as
\begin{equation*}%\label{eq:szr}
\resizebox{.9\textwidth}{!}{$
\begin{array}{rcccccccccc}
        &   & \text{\footnotesize Isolation} && \text{\footnotesize Exposure} && \text{\footnotesize Quarantine} && \text{\footnotesize Zombified} && \text{\footnotesize Removal}\\
\dot{S} & = & -\iota SZ + \kappa I & - & \epsilon SZ & & & & & & \\ %susceptible
\dot{I} & = & \iota SZ - \kappa I & & & & & & & & \\ %isolated
\dot{E} & = & & & \epsilon SZ & - & \omega E & - & \zeta_E E & - & \rho_E SE \\ %exposed
\dot{Q} & = & & & & & \omega E & - & \zeta_Q Q & - & \left(\rho_Q S + \delta\right)Q \\ %quarantined
\dot{Z} & = & & & & & \zeta_Q Q & + & \zeta_E E & - & \left(\rho_Z (S+I) + \delta\right)Z \\ %zombified
\dot{R} & = & & & & & & & & & \left(\rho_E E + \rho_Q Q + \rho_Z Z\right)S + \rho_Z IZ + \delta \left(Q + Z\right) %removed
\end{array}
$}
\end{equation*}
Briefly, this model encodes dynamics so that susceptible individuals enter isolation at a rate proportional to the susceptible and zombified populations ($\iota SZ$), but leave isolation based either purely on the boredom of the individuals or the idiocy of a fellow group member attracting zombies to the hideout ($\kappa I$).  Exposure occurs proportionally to the susceptible and zombified populations based on chance encounters ($\epsilon SZ$); exposed individuals can sacrifice themselves based on the surviving susceptible population ($\rho_E SE$), go into quarantine ($\omega E$), or become zombies ($\zeta_E E$).  Those exposed individuals in quarantine can either escape as zombies ($\zeta_Q Q$), be neutralized by the surviving population ($\rho_Q SQ$), or decompose through the natural process after death ($\delta Q$).  Finally, the zombified population can be be eradicated by the surviving population ($\rho_Z SZ$) using specialized tools, e.g., the lobo~\cite{worldwarz}, by the isolated community sniping from rooftops ($\rho_Z IZ$), or by the natural decomposition of the human body after death ($\delta Z$).  Notably, we ignore the risks of an increased death rate from isolated bands of humans coming into contact with each other (see, e.g., the events during the recent extended outbreak~\cite{walkingdead}\footnote{Be wary of those too fond of barbed wire-coated baseball bats named Lucille.}). 

Though a simpler model than that of~\cite{alemi2015zombies}, we utilize the calibration taken in that model (from studies of the outbreaks in~\cite{nightlivingdead,shaundead}) in order to determine many of our epidemiological factors.  We adjust these parameters by an order of magnitude to account for more recent outbreaks (see, e.g.,~\cite{walkingdead}) which can seemingly plod along relentlessly rather than the quick outbreaks studied previously in the literature.  Specifically, we find
\begin{equation*}
\begin{array}{rclcrclcrcl}
\epsilon & = & 3.60 \times 10^{-3}/pd & \qquad & & & & \qquad & & & \\
\zeta_E & = & 1.39/d & \qquad & \zeta_Q & = & 1.79 \times 10^{-3}/d & \qquad & & & \\
\rho_E & = & 0/pd & \qquad & \rho_Q & = & 0.90 \times 10^{-3}/pd & \qquad & \rho_Z & = & 2.88 \times 10^{-3}/pd\\
\delta & = & 82.24 \times 10^{-3}/d & \qquad & & & & \qquad & & &
\end{array}
\end{equation*}
with $d$ denoting days and $pd$ denoting person-days.  As the latent time for the zombie plague is quite short (corresponding to a half-life of 12 hours), we set $\rho_E = 0$ since, as observed in many prior outbreaks, susceptible individuals are generally unwilling to eliminate the threat from their soon-to-be zombified friends.  Additionally, we set $\delta = 82.24 \times 10^{-3}/d$ so as to account for the natural decomposition process of the human body~\cite{decomposition}, i.e., if you are zombified at time $t = 0$ you have a 90\% chance of decomposing 28 days later.\footnote{Pun intended.}  Notably, we leave the isolation ($\iota,\kappa$) and quarantine ($\omega$) parameters uncalibrated as these would be in response to policy implemented to encourage good public health standards and are thus variables for consideration later in this work.

\begin{figure}[t]
\begin{subfigure}[t]{0.48\textwidth}
\centering
\includegraphics[width=\textwidth]{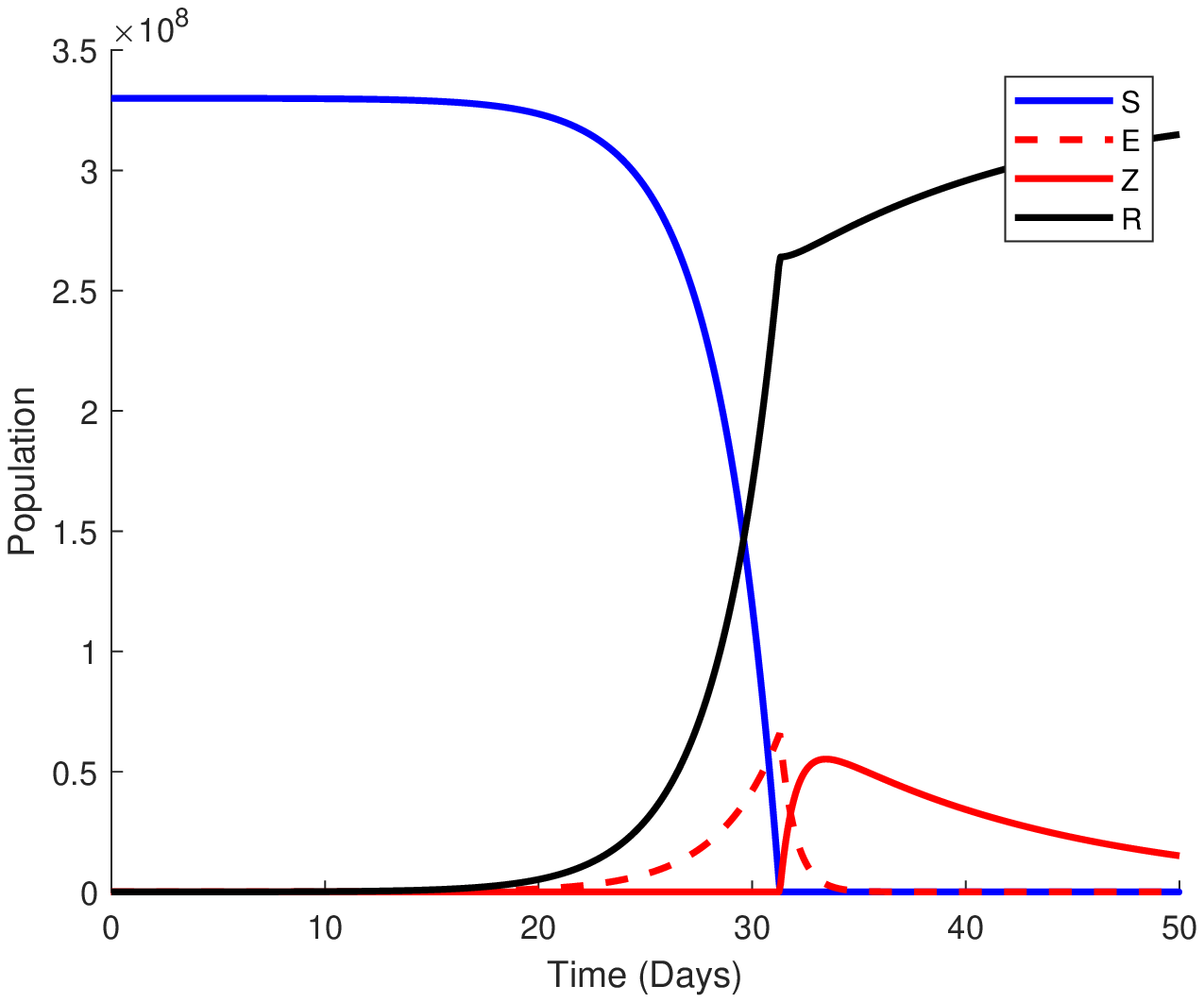}
\caption{Time evolution of the population during a zombie outbreak without isolation or quarantining.}
\label{fig:szr0}
\end{subfigure}
~
\begin{subfigure}[t]{0.48\textwidth}
\centering
\includegraphics[width=\textwidth]{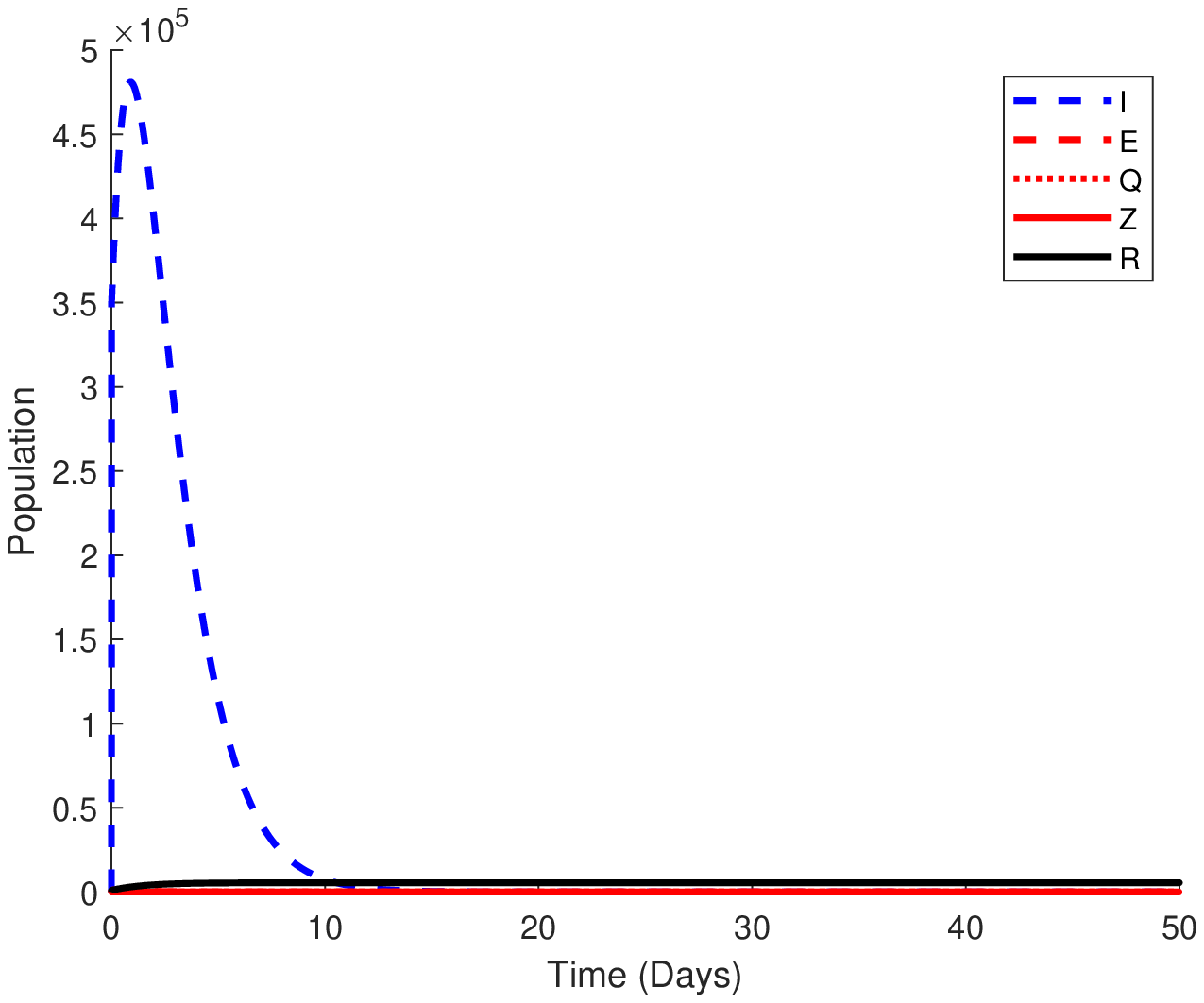}
\caption{Time evolution of the population during a zombie outbreak with strong isolation and quarantining policies.}
\label{fig:szr1}
\end{subfigure}
\caption{Sample epidemiological outputs of the SZR model considered.}
\label{fig:szr01}
\end{figure}

Prior works~\cite{munz2009zombies} primarily focus on the epidemiology of the zombie plague have considered the \emph{basic reproduction number $R_0$} of the outbreak and thus consider the likelihood of humanity's survival.  However, as humanity has continued through a multitude of zombie outbreaks in the modern era (see, e.g.,~\cite{nightlivingdead,dawnofdead1,dawnofdead2,shaundead,worldwarz,zombieland,walkingdead,worldwarz2}), we leave such considerations as an exercise to the reader.
We conclude this section with a demonstration of the lethality of the zombie epidemic without any isolation or quarantining.  That is, we set $\iota = \omega = 0$ (with $\kappa$ arbitrary).  Notably of the original 330 \emph{million} citizens of our sample major industrialized nation, %\footnote{Compare with 2018 population statistics in the United States \url{https://fred.stlouisfed.org/series/POPTOTUSA647NWDB}.}
 every single one would have been infected by the zombie plague in just over a single month.  This occurs even though the human population is able to contain, but not totally eradicate, the zombified population during this time.  The time evolution of the population is seen in Figure~\ref{fig:szr0}.  In contrast, a swift and strong isolation and quarantining policy $\iota = \omega = 1$ (with $\kappa = 0.693$ chosen so that the half-life of isolation is only a single day) completely eradicates the zombie epidemic without requiring significant isolation or quarantining of the populace as depicted in Figure~\ref{fig:szr1}.  Therefore, effective public response to a zombie crisis can not only flatten the curve, but entirely eliminate the outbreak before it gains a foothold.

\section{Economic and Financial Impacts}\label{sec:impacts}
\subsection{First-Order Effects on GDP}\label{sec:foe}
Now that we have a model for the propagation of the zombie plague through the population of a major industrialized nation, we wish to model the impacts these infections have on the economy and financial system given that a zombie outbreak has begun.  First, we will assume that isolated persons [I] completely drop off the grid in order to avoid the potential zombie horde.  As such, they neither produce nor purchase any goods or services during the period of self-imposed isolation.  Second, despite limited success getting zombified persons back to work in 2004~\cite{shaundead}, we assume all quarantined [Q] and zombified individuals [Z] drop out of the economy entirely.   Finally, and logically, any removed individuals [R] are now deceased and as such do not contribute to the economy.

The first-order effects of the zombie plague can thus be computed as the lost productivity from individuals who drop out of the economy.  Consider Gross Domestic Product [GDP] of \$21.73 \emph{trillion} for our major industrialized nation. %\footnote{Compare with 2019Q4 GDP for the United States \url{https://fred.stlouisfed.org/series/GDP}.}  
  With a population of 330 \emph{million}, we find a GDP/capita of roughly \$65,848.48 over a full year.  Furthermore, consider baseline consensus of 2\% growth in GDP over the next year; thus the losses are not in absolute terms, but measured against that metric.  Over the following year, the accumulated losses to potential GDP growth is computed as
\[\frac{1}{365}\int_0^{365} \left(I(t) + Q(t) + Z(t) + R(t)\right)dt \times \$65,848.48\]
where time is measured in days.  Of course, this metric is meaningless if the zombie outbreak becomes a true apocalypse and humanity is wiped out; for instance, in the setting in which no isolation or quarantining attempts are made there are losses of approximately \$19.995 \emph{trillion} over the first year after the outbreak; of course there are no humans remaining (and thus no economy) at the end of that time as depicted in Figure~\ref{fig:szr0} so this is, perhaps, irrelevant in such a scenario.

\subsection{Financial Systemic Risk}\label{sec:concept}
Provided a sufficient population of humans survive the zombie plague, the first-order effects do not tell the whole story.  An outbreak of the zombie plague is inherently unpredictable.  This uncertainty causes businesses and individuals to postpone major investments, decreasing demand for goods and services.  Additionally, hard-hit geographic areas both from the zombie plague and from policy responses cause individuals to withdraw from the workforce.  This can precipitate cascading effects because of supply chains; the goods produced by one person or company are the inputs for other goods and services in the larger economy.  For instance, the definitely-not-evil Umbrella Corporation manufacturers pharmaceuticals, which requires precise machinery and thousands of chemicals supplied by different companies. If any one of those companies fails to deliver the promised chemical or component, then the Umbrella Corporation may be forced to suspend production.  Such an event may cause the Umbrella Corporation to withdraw orders from all other suppliers causing cascading losses to the economy.

Conceptually, the zombie plague does nothing to impact debt obligations (e.g., mortgage or rent payments as well as bond offerings from corporations) even though both the supply and demand of goods and services may dry up.  Long-term debt is not a concern as the zombie outbreak will either: decimate society so as to make John Maynard Keynes correct that ``in the long run we are \emph{all} dead;'' or eventually the zombie plague will be defeated and the economy will recover.  However, short-term obligations will still require payments and, furthermore, zombies have a poor record of repaying even their long-term obligations.  If short-term lending markets remain liquid (i.e., willing and able to provide the necessary funds) then there may only be limited impacts, as the financial system will functionally transform those short-term debts to longer-term obligations (with some additional interest payments).  If the short-term lending markets become illiquid due to uncertainty about future payments (e.g., financing would not be available to any individual or business that may soon be zombified), then individuals and companies alike will need to raise cash in other ways.  This deleveraging to reduce debt would be accomplished by selling goods (such as collectibles or financial assets), but, with a deluge of assets sold at once, the price of each item will drop.  Such an event is often called a fire sale; importantly, when fire sales are triggered by frozen short-term lending markets (rather than, e.g., from panic associated with the popping of an economic or financial bubble) the typical response of lowering interest rates will have no effect on the speed and severity of asset liquidations, since the interest rates only alter the cost at which borrowing occurs.  If this causes financial securities (e.g., stocks and bonds) to fall too quickly (potentially coupled with writedowns on loans to zombified individuals), this can cause the banks themselves to need to sell the assets (in, e.g., a margin call) they hold which has a dual impact: (1) the value of the financial markets will plummet in what is called price-mediated contagion and (2) new (short-term) cash will also become more difficult to obtain further exacerbating the fire sale.  This cyclical deleveraging is one example of, and often called, \emph{systemic risk}.

If you, the reader, are afraid of mathematical details enriching your brain, thereby attracting zombies, proceed to Section~\ref{sec:results} without any loss of conceptual understanding. 

\subsection{Modeling Details}\label{sec:details}
With this conceptual understanding, we now wish to investigate the scale of the systemic risk.  We will begin with a consideration of household debts and the impact that the zombie plague may have on the repayment of those obligations.  In the major industrialized nation under consideration, current outstanding household debt stands at \$16.58 \emph{trillion}, %\footnote{Compare with 2019Q4 household liabilities in the United States \url{https://www.federalreserve.gov/releases/z1/dataviz/z1/balance\_sheet/table/}.} 
 which requires approximately 15.03\% of after-tax income for debt payments (i.e., 15.03\% of \$68,527 per year per household). %\footnote{Compare with 2019Q3 household financial obligations as a percent of disposable personal income in the United States \url{https://fred.stlouisfed.org/series/FODSP}.}
  Thus the average household (of 2.52 persons) has \$10,299.64 of debt payments per year for a population total of \$1.35 \emph{trillion}.  Thus the writedowns on household debt from zombified, quarantined, and removed individuals $T$ days after the outbreak began should be 
\[
\resizebox{.9\textwidth}{!}{$
\int_0^{T} \left(Q(t) + Z(t) + R(t)\right) dt \times \frac{\$1.35 \times 10^{12}}{330 \times 10^{6}} + \left(Q(T) + Z(T) + R(T)\right) \times \left(\frac{\$16.58 \times 10^{12}}{330 \times 10^{6}} - \frac{\$1.35 \times 10^{12} \times T}{330 \times 10^{6} \times 365}\right).
$}
\]
However, during the outbreak itself, lenders may not have reliable data on who has been zombified and thus we will assume that these markdowns do not occur until after the zombie plague has been defeated once more.
Additionally, individuals in isolation will lose out on income during their time of seclusion.  To simplify, if short-term lending is available then all individuals take advantage with minimal cost; without short-term lending available, we assume all lost income is compensated for by selling assets, i.e., isolated individuals sell a dollar value of
\[\int_0^T I(t)dt \times \frac{\$51,428}{365}\]
from time $0$ to $T$ days after the outbreak based on an average after-tax income of \$51,428 (i.e., approximately 1.33 earners per household).  To simplify our setting, we assume that all individuals hold a diversified portfolio which is liquidated proportional to the holdings and, importantly, have enough investments to cover these short-term losses.

In addition to the household debts, businesses in our major industrialized nation have obligations totalling \$29.45 \emph{trillion}. %\footnote{Compare with 2018 private debt as percent of GDP for the United States \url{https://www.imf.org/external/datamapper/Privatedebt\_all@GDD/CAN/GBR/DEU/ITA/FRA/JPN/USA}.}
  In fact, even companies can catch the zombie plague~\cite{banerjee2018rise}, but to simplify considerations we ignore the very real threats from zombie firms unable to pay their debts.
  With a debt service coverage ratio of 1.35 and yearly operating income of \$8.25 \emph{trillion} results in a debt payment of \$6.11 \emph{trillion} within the next year. %\footnote{Compare with corporate profits in the United States \url{https://www.statista.com/statistics/222127/quarterly-corporate-profits-in-the-us/}.}
  In order to maintain this debt service coverage ratio, as is typically mandated by lenders (and reducing investments and other expenditures to make up the short-term debt servicing), the businesses will need to make up for lost profits (of \$25,000 per person per year) by liquidating assets as well.  The dollar value of these liquidations up to day $T$ is provided by
\begin{align*}
\frac{1}{1.35} \times &\left[\int_0^T \left(I(t) + Q(t) + Z(t) + R(t)\right)dt \times \$25,000 \right.\\
&\qquad\left. + \left(Q(T) + Z(T) + R(T)\right) \times \left(\$120,477-\frac{\$25,000 \times T}{365}\right)\right].
\end{align*}
As with households, we assume these businesses sell the market as a whole as necessary. 

With these results, we can now consider the impacts of all of these asset liquidations and writedowns on the financial sector and markets.  With a well-functioning repurchase agreement [repo] market, all banks remain well capitalized in the short term and thus can continue to lend to individuals and businesses.  While some asset liquidations would occur, as shown in~\cite{bichuch2018borrowing}, there are only minor price impacts and no major systemic risk event occurs.  However, without a functioning repo market, the banks are themselves unable to obtain short term funding in order to satisfy margin calls on existing leveraged agreements which are proportional to the risk-weighted assets at a level 25\% above the regulatory threshold (with $\theta_{\min} = 0.1$ and with risk-weights $\alpha$ and price impacts $b$ defined in Table~\ref{tab:rw}).  Specifically, we consider 6 systemically important financial institutions for our major industrialized nation described and calibrated in~\cite{BW19,BF19}.  Notably, these banks do \emph{not} invest solely in the market portfolio but do so strategically.  Under the assumption that banks sell the minimum necessary to satisfy the margin requirements, and following a proportional liquidation strategy as in~\cite{GLT15,DE18}, the price-mediated contagion during the zombie outbreak is described as in~\cite{feinstein2019leverage}, i.e., the proportion of asset liquidations $\Pi$, asset prices $q$, value of cash accumulated from liquidations $\Psi$, and proportion of liquidations by the non-banking economy $\eta$ follow the dynamics
\begin{align*}
\dot \Pi &= -\left(I + \Lambda(t) \diag[-b]\diag[q] s^\T\right)^{-1} \Lambda(t) \diag[-b]\diag[q]M\dot{\eta}\\
\dot q &= \left(I + \diag[-b]\diag[q] s^\T \Lambda(t)\right)^{-1} \diag[-b]\diag[q]M\dot{\eta}\\
\dot \Psi &= \diag[\dot \Pi] s q\\
\Lambda(t) &= \diag\left[\ind{\theta \leq \theta_{\min}}\right] \diag\left[s A q \theta_{\min}\right]^{-1} (s - \Gamma) (I - A\theta_{\min})\\
\dot{\eta} &= \frac{1}{M^\T q} \left(\frac{\$51,428}{365} \times I \right.\\
    &\quad\quad \left.+\frac{1}{1.35}\left[\$25,000 \times \left(I + \frac{364}{365}\left(Q + Z + R\right)\right) + \left(\$120,477 - \frac{\$25,000 \times t}{365}\right)\times\left(\dot Q + \dot Z + \dot R\right)\right]\right)
\end{align*}
where $\theta_i \leq \theta_{\min}$ determines if a bank is subject to a margin call and needs to sell assets at all, $A := \diag[\alpha]$ for notational simplicity, $M$ is the market capitalization for each asset (\$30.42 \emph{trillion} divided amongst the 16 assets in same proportion to the banking sector as a whole), and $s$ is the matrix of bank investments.  Once the zombie plague has been eradicated, each bank takes stock of its loan portfolio ($\ell$) and marks down any obligations from those that had been zombified.  This is captured by adjusting the loan portfolios downward as computed above and the subsequent fire sale is modeled as in~\cite{BF19}, i.e., $q_*$ is the final market prices and $\mu$ is the vector of writedowns for each bank
\begin{align*}
q_* &= \diag[q(T)]\exp(-\diag[b]\sum_{i = 1}^6 \gamma_i)\\
\bar q_{*,k} &= \frac{q_k(T)}{b_k \sum_{i = 1}^6 \gamma_{ik}} \left[1 - \exp\left(-b_k \sum_{i = 1}^6 \gamma_{ik}\right)\right]\\
\gamma_i &= \left[\left(\frac{h_i - q_*^\T[I - A\theta_{\min}](1-\Pi_i(T))s_i}{(\bar q_* - [I - A\theta_{\min}]q_*)^\T (1-\Pi_i(T))s_i}\right)^+ \wedge 1\right](1 - \Pi_i(T))s_i\\
h &= \bar p - x - \Psi(T) - \diag[1 - \alpha_{\ell}\theta_{\min}](\ell - \mu)\\
\mu &= \frac{\ell}{\sum_{i = 1}^6 \ell_i} \left[\int_0^{T} \left(Q(t) + Z(t) + R(t)\right) dt \times \frac{\$1.35 \times 10^{12}}{330 \times 10^{6}} \right. \\ 
    &\quad\quad \left. + \left(Q(T) + Z(T) + R(T)\right) \times \left(\frac{\$16.58 \times 10^{12}}{330 \times 10^{6}} - \frac{\$1.35 \times 10^{12} \times T}{330 \times 10^{6} \times 365}\right)\right]
\end{align*}
\begin{table}[t]
\centering
{\footnotesize
\begin{tabular}{|c|cccccccc|}
\hline
\textbf{Asset} & 1    & 2    & 3   & 4    & 5    & 6    & 7   & 8  \\ \hline %  & 9    & 10  & 11  & 12   & 13 & 14  & 15   & 16  \\ \hline
$\alpha$ & 0.07 & 0.08 & 0.1 & 0.12 & 0.15 & 0.18 & 0.2 & 0.25 \\ \hline %& 0.35 & 0.5 & 0.6 & 0.75 & 1  & 2.5 & 4.25 & 6.5 \\ \hline
%$M$ & 256.8769 & 256.2603 & 255.0269 & 253.7936 & 251.9435 & 250.0935 & 248.8602 & 245.7768 & 239.6101 & 230.3600 & 224.1932 & 214.9431 & 199.5263 & 107.0253 & 36.3812 & 9.4691 \\ \hline
%$b$ ($10^{-15}$) & 3.61 & 4.13 & 5.17 & 6.22 & 7.80 & 9.38 & 10.45 & 13.13 & 18.57 & 26.95 & 32.68 & 41.51 & 56.89 & 170.67 & 378.43 & 950.86 \\ \hline 
$b$ ($10^{-23}$) & 2.20 & 2.52 & 3.17 & 3.83 & 4.84 & 5.86 & 6.56 & 8.35 \\ \hline\hline %& 12.11 & 18.28 & 22.78 & 30.18 & 44.55 & 249.16 & 1,625.30 & 15,690.21 \\ \hline
\textbf{Asset} & 9 & 10 & 11 & 12 & 13 & 14 & 15 & 16 \\ \hline
$\alpha$ & 0.35 & 0.5 & 0.6 & 0.75 & 1  & 2.5 & 4.25 & 6.5 \\ \hline
$b$ ($10^{-23}$) & 12.11 & 18.28 & 22.78 & 30.18 & 44.55 & 249.16 & 1,625.30 & 15,690.21 \\ \hline
\end{tabular}\vspace{1em}
\begin{tabular}{|c|cccccc|}
\hline
\textbf{Bank} & BAC & C & GS & JPM & MS & WFC \\ \hline
$\alpha_{\ell}$ & 0.8464 & 0.8977 & 0.7222 & 0.7735 & 0.7212 & 0.8503 \\ \hline
\end{tabular}
}
\caption{Risk-weights $\alpha$ and market impacts $b$ for the marketable assets and banks.}
\label{tab:rw}
\end{table}

\section{Public Policy and Impacts on Financial Markets}\label{sec:results}
We wish to conclude this study by considering the impacts of different isolation and quarantine strategies on survival and financial markets.  As shown in Figure~\ref{fig:szr01}, having no isolation or quarantine strategy leads to the complete annihilation of humanity, whereas a robust isolation and quarantine strategy can entirely eradicate the zombie plague in a matter of days.  
Throughout this section we set $\kappa = 0.0495$ so that the median isolated individual goes into hiding for a fortnight.  Additionally, as previous outbreaks have taught us that bitten individuals tend to act extremely selfishly and tend to ignore their own risks, we consider $\iota = \omega$ so that exposed individuals enter an isolated quarantine at the same rate as the susceptible population enters isolation.

\begin{figure}[t]
\begin{subfigure}[t]{0.48\textwidth}
\centering
\includegraphics[width=\textwidth]{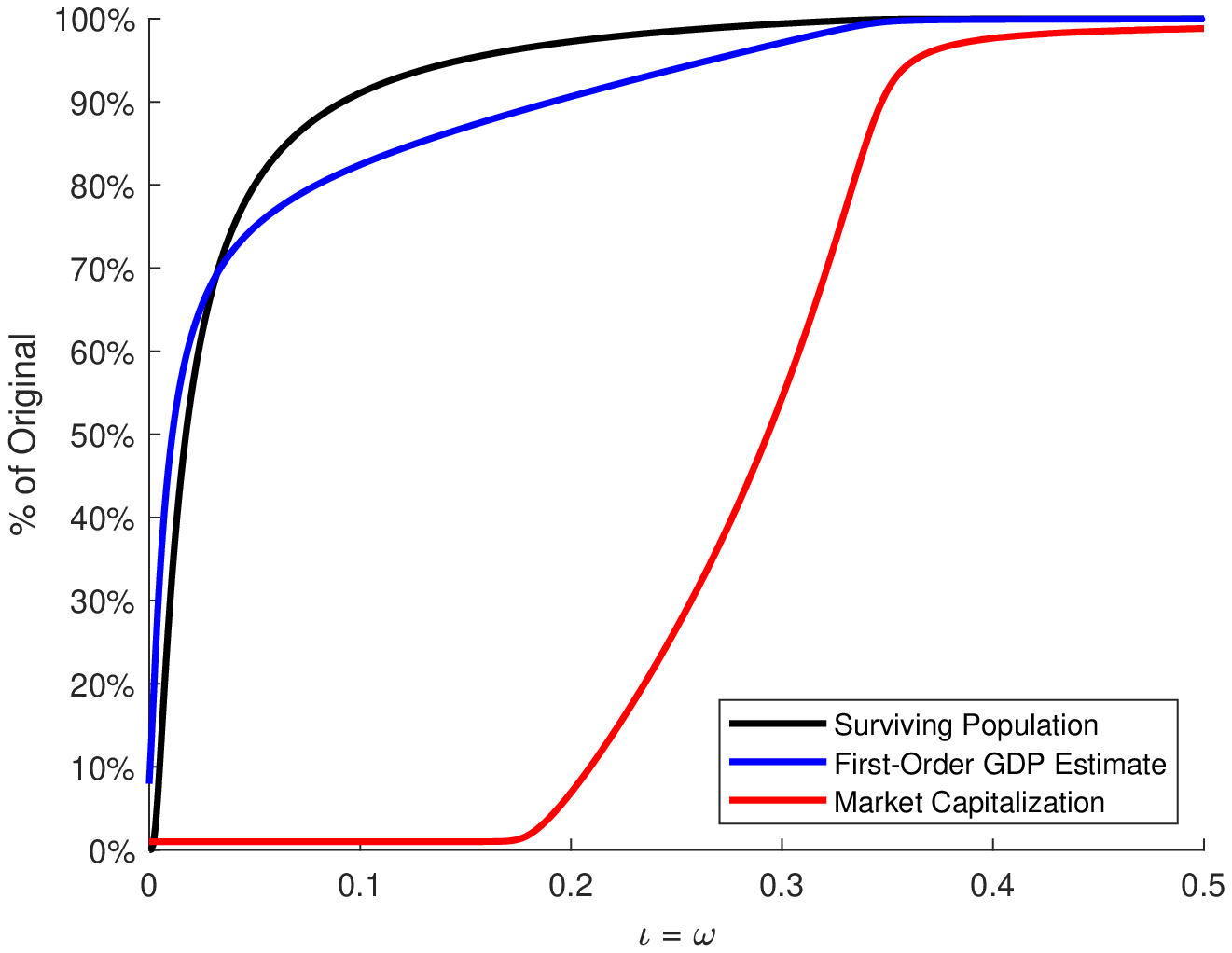}
\caption{Impact of isolation and quarantine policy on the survival of the human race and financial markets 1 year after the zombie outbreak.}
\label{fig:impact1}
\end{subfigure}
~
\begin{subfigure}[t]{0.48\textwidth}
\centering
\includegraphics[width=\textwidth]{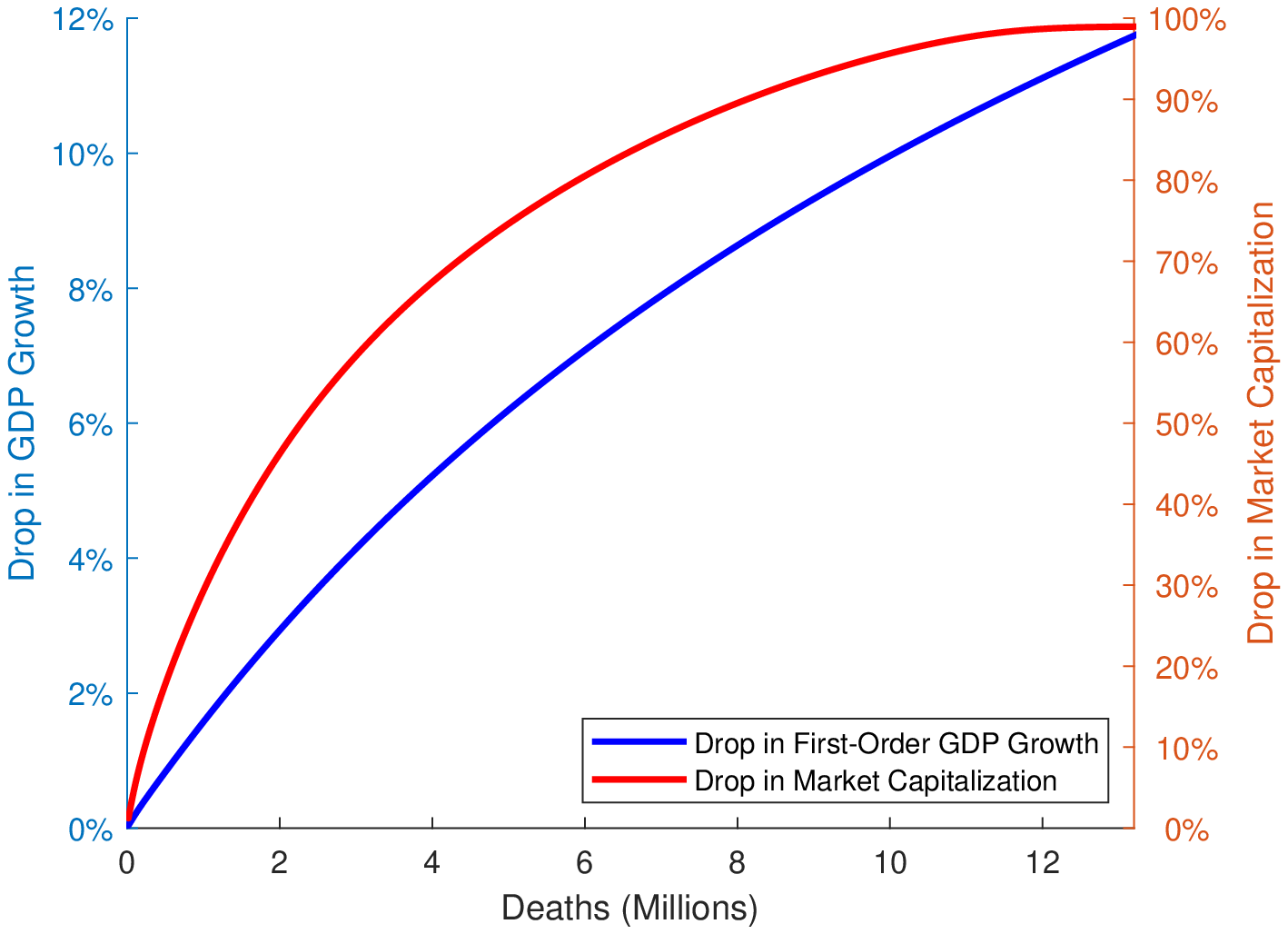}
\caption{Relationship between deaths caused by the zombie plague, GDP losses, and the drawdowns in financial markets 1 year after the zombie outbreak.}
\label{fig:impact2}
\end{subfigure}
\caption{Impacts of public health policy on survival, GDP, and financial markets.}
\label{fig:impact}
\end{figure}

The impacts of varying $\iota = \omega$ on the survival of the human species and on market capitalization of the full financial market after a single year is displayed in Figure~\ref{fig:impact}.  As with some other pandemics~\cite{coronavirus}, taking multiple simultaneous isolation strategies can greatly reduce the severity and mortality of the zombie plague, though not eliminate it entirely.  Recall that with a population of 330 \emph{million}, even 1\% mortality of the general populace in our major industrialized nation is 3.3 \emph{million} deaths.  However, the economy and financial markets are not so forgiving of millions of deaths.  For instance, based on these simulations, a zombie outbreak that kills 1 \emph{million} people leads to first-order GDP losses (from baseline growth model of, e.g., 2\% growth) of 1.57\% and a drop in the full market capitalization of 29.3\%; a summary of economic results is provided in Table~\ref{tab:economic}. %drops by 3.28\% and 50\%, respectively, in a zombie outbreak that kills off 2.28 \emph{million} people in our major industrialized nation.  
Recall that these first-order GDP losses are likely an underestimate as they only consider the lost productivity but no feedback effects.  
Thus the best economic and financial solution is to eliminate the zombie threat before it rises from the grave to threaten the world.

To capture some of the feedback effects on GDP, we consider the market capitalization-to-GDP ratio.  With a long-run average over the past 25 years of 125\%, any drop in the market capitalization can be expected to cause 80\% as much damage to GDP; we will call this the regressed GDP estimate.\footnote{Though the pre-outbreak ratio is 1.4, such a ratio tends to drop during market crashes. Due to this we consider the long-run average rather than the current ratio.} With such a regression, the GDP drops 23.44\% (in absolute terms) due to a zombie outbreak that kills 1 \emph{million} people.
%linear regression of GDP with respect to the performance of financial markets.\footnote{Due to data limitations, this result should not be considered reliable beyond a 50\% drop in the financial markets and as such is not included in the plots.}  With such an estimate, the GDP drops 4.27\% from the baseline due to a zombie outbreak that kills 1 \emph{million} people.  
\begin{table}[t]
\centering
\begin{tabular}{|c||c|c|c|c|c|}
\hline
%CHOOSE \iota = 0.5, 1 million deaths, 50\% drop in market, 4.3\% GDP (LOOK UP 2008!)
$\iota = \omega$ & 0.500 & 0.322 & 0.293 & 0.274 & 0.266 \\ \hline\hline %SORT HIGH TO LOW
Deaths (\emph{Millions}) & 0.016 & 1.000 & 2.280 & 3.300 & 3.787 \\ \hline %SORT LOW TO HIGH
First-Order GDP Losses (\%) & 0.04 & 1.57 & 3.28 & 4.48 & 5.00 \\ \hline
Regressed GDP Losses (\%) & 0.94 & 23.44 & 40.00 & 49.03 & 52.53 \\ \hline
Market Cap Losses (\%) & 1.18 & 29.30 & 50.00 & 61.29 & 65.66 \\ \hline
\end{tabular}
\caption{Summary of health, economic, and financial impacts of the zombie plague over 1 year for select public health strategies.}
\label{tab:economic}
\end{table}

Though obviously the best case scenario is to prevent and contain a zombie apocalypse, herein we recommend some strategies to mitigate the worst case scenarios from a zombie outbreak.  First, and foremost, listening to experts as to whether to self-isolate or prepare a secure quarantine zombie zone is vitally important.  As demonstrated in Figure~\ref{fig:szr1}, a strong policy of isolation and quarantine can eliminate a zombie outbreak in days with few deaths; as demonstrated in Figure~\ref{fig:impact}, the quickest way to rescue the financial markets is to prevent the zombie epidemic from spreading.  However, the resources to implement such a robust strategy are immense; for example, the closing of shopping malls and bars (such as the Monroeville Mall and Winchester Tavern) is costly.  Thus it is, perhaps, overly optimistic to believe such a strategy will be undertaken early against an outbreak of the zombie plague.  

Therefore, with the above considerations of cost of allowing the zombified population to grow, the government and central bank of our major industrialized nation should guarantee that short-term lending markets remain active to prevent a run on the shadow banking system.  A rescue of the repo market, along with incentives for debt restructuring for impacted individuals and businesses to avoid the vast majority of zombie-based defaults, would eliminate the greatest harms to the financial markets.  As such, in opportunity costs, this rescue strategy could pay for itself many times over.

\section{Conclusion}\label{sec:conclusion}
We modeled the state of the financial markets of a major industrialized nation from an outbreak of the zombie plague.  In order to accomplish such a feat, we consider the epidemiological modeling of the zombie plague through the population and then model how such contagion spreads to and through the financial system.  This modeling allowed us to consider how different public health policies on isolation and quarantining the population can mitigate the zombie plague on both the population and financial markets; in our nation of 330 \emph{million} people, a loss of 1 \emph{million} people to the zombie plague can cause GDP losses of 23.44\% and a market drop of 29.30\%.
The costs of ignoring the economics and financial impacts of the zombie plague can be apocalyptic, so following these mitigation strategies is paramount.  And remember, if you do not use your brain during a zombie outbreak, you may start craving someone else's.

{\footnotesize
\bibliographystyle{plain}
\bibliography{bibtex3}
}

\appendix
\section*{Glossary of Symbols}
\begin{tabular}{p{0.03\textwidth}p{\dimexpr 0.92\textwidth-2\tabcolsep}}
$\alpha$ & Vector of risk-weights for the financial assets\\
$\alpha_{\ell}$ & Vector of risk-weights for bank loans\\
$\gamma$ & Vector of bank liquidations due to writedowns in loans following the zombie outbreak\\
$\delta$ & Rate that zombies decompose\\
$\epsilon$ & Rate that susceptible individuals are bitten by zombies\\
$\zeta_E$ & Rate that exposed individuals turn into zombies\\
$\zeta_Q$ & Rate that quarantined zombies escape from their confinement\\
$\eta$ & Total market liquidations by non-financial sector during the zombie outbreak\\
$\theta_{\min}$ & Margin requirement for banks\\
$\iota$ & Rate that susceptible individuals go into isolation\\
$\kappa$ & Rate that isolated individuals leave their hiding places due to boredom or being found\\
$\rho_E$ & Rate that susceptible individuals eliminate the threat from their bitten friend\\
$\rho_Q$ & Rate that susceptible individuals find and neutralize quarantined zombies\\
$\rho_Z$ & Rate that susceptible individuals successfully double tap zombies\\
$\mu$ & Vector of bank losses to their loans due to writedowns due to the zombie outbreak\\
$\omega$ & Rate that exposed individuals quarantine themselves before becoming a zombie\\
$\Lambda$ & Financial contagion matrix\\
$\Pi$ & Vector of proportions of bank trading book liquidated during the zombie outbreak\\
$\Psi$ & Vector of cash raised by banks due to liquidations during the zombie outbreak\\
$b$ & Vector of market impact parameters for each asset\\
$h$ & Vector of bank shortfalls following the zombie outbreak\\
$\ell$ & Vector of loan portfolios for banks\\
$\bar p$ & Vector of liabilities for banks\\
$q$ & Vector of asset prices during the zombie outbreak\\
$q_*$ & Vector of asset prices following bank liquidations following the zombie outbreak\\
$\bar q_*$ & Vector of asset prices obtained by banks when liquidating following the zombie outbreak\\
$s$ & Matrix of bank asset holdings\\
$x$ & Vector of cash available to the banks\\
A & Diagonal matrix of risk-weights $\alpha$\\
E & Number of individuals exposed to the zombie plague\\
I & Number of individuals practicing self-isolation\\
M & Vector of asset market capitalizations\\
Q & Number of bitten individuals in quarantine from the general populace\\
R & Number of individuals removed from the system so they are no longer a threat\\
S & Number of individuals susceptible to the zombie plague\\
Z & Number of zombies
\end{tabular}

\end{document}